\documentclass[a4paper,11pt]{article}
\pdfoutput=1 

\usepackage{jcappub} 

\usepackage[T1]{fontenc} 
\usepackage{soul}
\graphicspath{{./figs/}}

\newcommand{\lcdm}{$\Lambda{\rm CDM}$}

\title{\boldmath The angular scale of homogeneity with SDSS-IV DR16 Luminous Red Galaxies}

\author[a]{Uendert Andrade,}
\author[a,b]{Rodrigo S. Gon\c{c}alves,}
\author[c]{Gabriela C. Carvalho,}
\author[a]{Carlos A. P. Bengaly,}
\author[a]{Joel C. Carvalho}
\author[a]{and Jailson Alcaniz}

\affiliation[a]{Observat\'orio Nacional, 20921-400, Rio de Janeiro, RJ, Brazil}
\affiliation[b]{Departamento de F\'isica, Universidade Federal Rural do Rio de Janeiro, Serop\'edica, Rio de Janeiro 23897-000, Brazil}
\affiliation[c]{Universidade do Estado do Rio de Janeiro, Faculdade de Tecnologia, 27537-000, Resende - RJ, Brazil}

\emailAdd{uendertandrade@on.br}
\emailAdd{rsg\underline{ }goncalves@ufrrj.br}
\emailAdd{gabriela.coutinho@fat.uerj.br}
\emailAdd{carlosbengaly@on.br}
\emailAdd{jcarvalho@on.br}
\emailAdd{alcaniz@on.br}

\abstract{We report measurements of the angular scale of cosmic homogeneity ($\theta_{H}$) using the recently released luminous red galaxy sample of the sixteenth data release of the Sloan Digital Sky Survey (SDSS-IV LRG DR16). It consists of a model-independent method, as we only use the celestial coordinates of these objects to carry out such an analysis. The observational data is divided into thin redshift bins, namely $0.67<z<0.68$, $0.70<z<0.71$, and $0.73<z<0.74$, in order to avoid projection biases, and we estimate our uncertainties through a bootstrap method and a suite of mock catalogues. We find that the LRGs exhibit an angular scale of homogeneity consistent with the predictions of the standard cosmology within the redshift interval studied. Considering the bootstrap method, in which the measurements are obtained in a model-independent way, we found at 1$\sigma$ level that $\theta_H^{boot}(0.675) = 7.57 \pm 2.91$ deg, $\theta_H^{boot} (0.705) = 7.49 \pm 2.63$ deg and $\theta_H^{boot} (0.735) = 8.88 \pm 2.81$ deg. Such results are in good agreement with the ones obtained using mock catalogues built under the assumption of the standard cosmological model.}

\begin{document}
\maketitle
\flushbottom

\section{Introduction}
\label{sec:intro}

\indent
The Cosmological Principle (CP) is one of the most fundamental pillars of modern cosmology. Given the success of the $\Lambda$CDM cosmology in describing the observed Universe~\cite{planck18, pantheon18, sdss21, kids21, des21a, des21b}, which assumes large-scale homogeneity and isotropy with structure formation described via perturbations, the CP has been indirectly established as a valid assumption, but still rarely tested directly through cosmological observations. Although tests of the isotropy hypothesis have been performed in a direct manner, homogeneity is much harder to probe by observations, since source counts on spatial hypersurfaces inside the past lightcone cannot be accessed by this method - in fact, these counts are limited to the intersection of the past lightcone with the spatial hypersurfaces (see, e.g., ~\cite{Clarkson:2010uz, Maartens:2011yx, Clarkson:2012bg} for a broad discussion). 

It is well known that small-scale inhomogeneities are expected in a Friedmann-Lema\^itre-Robertson-Walker (FLRW) Universe, so that we observe a complex web of structures composed of non-uniformities such as groups and clusters of galaxies, voids, walls, and filaments. In such a background, we also expect a transition scale from a lumpy to a smoother, more homogeneous Universe, above which the patterns created by these structures become indistinguishable from a random distribution of sources. This transition scale corresponds to the so-called \emph{cosmic homogeneity scale}, $r_{\rm h}$, which has been identified and measured by several galaxy and quasar surveys at around $70 - 150 \, \mathrm{Mpc}$~\cite{Hogg:2004vw, Sarkar:2009, Scrimgeour:2012wt, Pandey:2013, Pandey:2015xea, Pandey:2016, Laurent:2016eqo, Ntelis:2017nrj, Goncalves:2018sxa, Goncalves:2020erb, Avila:2021mbj, Kim:2021osl}. This is in excellent agreement with the upper level of $r_{\rm h} \simeq 260 \, \mathrm{Mpc}$ predicted by~\cite{Yadav:2010cc}. These results have been confirmed by analyses relying on different methods and data~\cite{Zhang:2010fa, Hoyle:2012pb, Valkenburg:2012td, Kraljic:2015, Jimenez:2019cll, Camarena:2021mjr}, although some authors have claimed the absence of such homogeneity scale~\cite{Labini:2009ke, Labini:2011dv, Park:2016xfp}, arguing that these measurements could be biased by the survey window function~\cite{Heinesen:2020wai}. Moreover, some tests of cosmic isotropy revealed potential hints at FLRW breakdown ~\cite{Bengaly:2017slg, Colin:2019opb,Migkas:2020fza, Secrest:2020has, Krishnan:2021jmh, Singal:2021crs, Singal:2021kuu, Luongo:2021nqh}, albeit disputed by~\cite{Bengaly:2017zlo, Andrade:2018eta, Andrade:2019kvl, Lukovic:2019ryg, Akrami:2019bkn, Kazantzidis:2020tko, Rahman:2021mti}. In light of these results, it is crucial to perform new tests of the CP in order to determine whether it constitutes a valid physical assumption. If otherwise, a complete reformulation of the standard cosmological paradigm would have to be pursued~\cite{Ellis:2006fy}. 

In this work, we carry out a test of the statistical homogeneity of the Universe using the Luminous Red Galaxies (LRG) catalogue by the sixteenth release of SDSS-IV~\cite{DR16Q, eboss20, eboss21}. Rather than measuring the three-dimensional homogeneity scale $r_{\rm h}$, as in previous works~\cite{Goncalves:2018sxa, Goncalves:2020erb}, we focus on the \emph{angular scale of homogeneity}. This analysis circumvents the necessity to convert the redshifts into cosmological distances, which relies on the assumption of a cosmological model. Hence, measuring the angular homogeneity constitutes a \emph{model-independent test} of the cosmic homogeneity hypothesis, because we only depend on the celestial coordinates of each object. Such a test was originally proposed by~\cite{Alonso:2013boa}, and performed by~\cite{Alonso:2014xca, Goncalves:2017dzs, Avila:2019gdb, Camacho2021:sdf} using observational data from the 2MASS photo-z catalogue, SDSS-III DR12 LRGs, SDSS blue galaxies, and Planck's CMB temperature anisotropy maps, respectively. The first works showed that the angular homogeneity scale measurements are in good agreement with the standard model expectations, albeit the concordance was weaker in~\cite{Camacho2021:sdf}. In addition, our analysis probes higher redshift ranges than previous galaxy-based works, since the latest SDSS LRG dataset is composed of objects in the $0.6<z<1.0$ interval. Therefore, we are able to verify if the angular homogeneity scale can be identified in such epochs of the Universe, and how it evolves with redshift. 

Another source of model-dependency usually comes from mock catalogues needed to estimate the covariance matrix. In fact, simulations are seeded using external cosmological information besides the data itself. This type of strategy has been coined as external methods and differs from internal methods that use only the data to obtain the covariance matrix~\cite{Norberg:2008tg}. Here, we estimate the uncertainties by means of the bootstrap method (internal) and compare our results with 1000 mock catalogues produced under the assumption of the standard cosmological model (external). Thus, we can not only estimate our uncertainties in a model-independent way, but also verify whether the standard model predictions for the angular homogeneity scale are consistent with the observational measurements.

We organize this paper as follows: Section 2 describes the observational sample used in our analysis; Section 3 is dedicated to the method developed to measure the angular homogeneity scale, and the theoretical expectations; Section 4 shows our results and discussions; Concluding remarks are presented in Section 5.


\section{Observational Data}
\label{sec:ObsData}

\indent

\begin{table}[b!]
\centering
\begin{tabular}{|c|c|c|}
\hline
\,\,$z$ \,\,& \,\, $\bar{z}$ \,\,& \,\,$N_{\rm galaxies}$\,\, \\
\hline
\,\,\,0.67-0.68 \,\,& 0.675 & 4209 \\
\,\,\,0.70-0.71 \,\,& 0.705 & 4073 \\
\,\,\,0.73-0.74 \,\,& 0.735 & 4210 \\
\hline
\end{tabular}
\caption{The redshift bins adopted in our analysis, along with the redshit bin means and corresponding number of LRGs in each bin.}
\label{table1}
\end{table}

\begin{figure}[t!]
	\centering
	\includegraphics[scale=0.52]{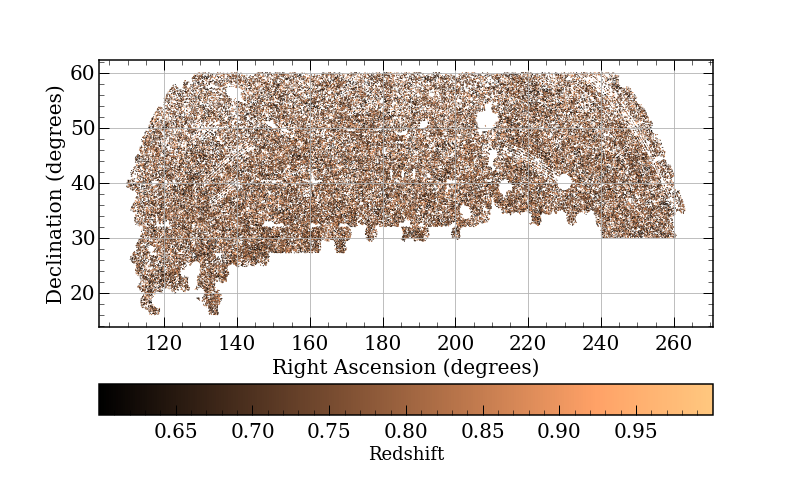}
    \caption{The footprint of SDSS-IV LRG DR16 catalogue.}
    \label{fig:fig1}
\end{figure}

The final public data of the fourth phase of the Sloan Digital Sky Survey (DR16) provides the optical spectra of Luminous Red Galaxies (LRGs), including those designed to test target selection algorithms for eBOSS, called the Sloan Extended Quasar, ELG, and LRG Survey (SEQUELS). We use the latest LRG catalogue provided by the SDSS-IV collaboration~\cite{eboss20}, namely SDSS LRG DR16, which contains $174,816$ LRGs within a redshift interval of $0.6<z<1.0$ covering a total area of the sky of $4,242 \; \mathrm{deg}^2$. This sample is divided into two hemispheres, namely the North Galactic Cap (NGC) and the South Galactic Cap (SGC). We restrict our analysis to the former due to the limited sky coverage of the latter, ending up with $107,500$ objects within a sky area of $2,566  \; \mathrm{deg}^2$, as displayed in~Fig.\ref{fig:fig1}. 

\begin{figure}[t!]
	\centering
	\includegraphics[scale=0.52]{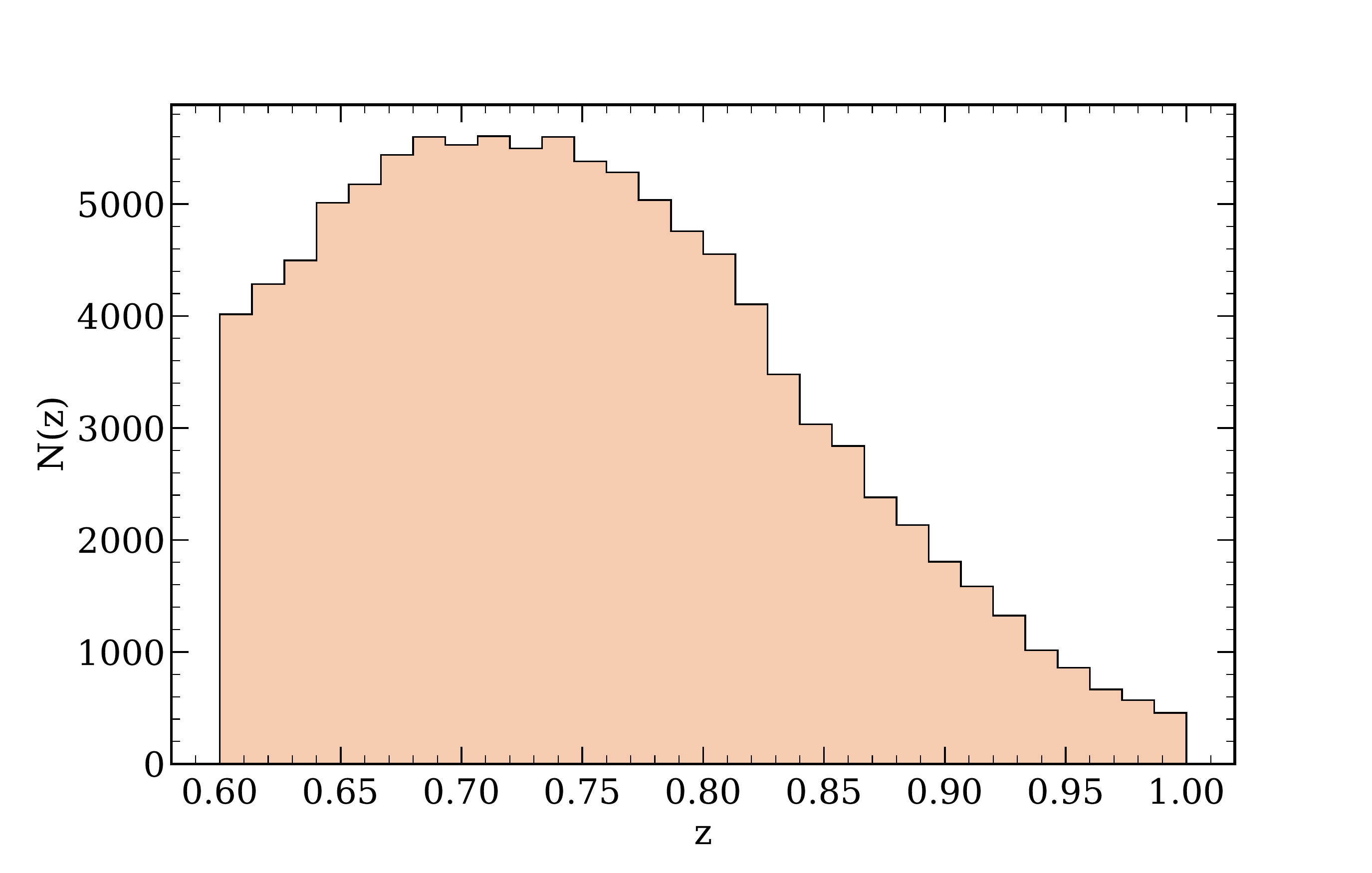}
    \caption{Redshift distribution of the luminous red galaxies
for SDSS-IV DR16. We considered the redshift interval $0.67<z<0.68$, $0.70<z<0.71$, and $0.73<z<0.74$ in our analysis as these intervals are closer to the distribution peak.}
    \label{fig:fig2}
\end{figure}

As we focus on measuring the angular scale of homogeneity, we split the data into thin redshift bins, whose main features are shown in Table~\ref{table1}. Each bin has about $~\mathcal{O}(10^3)$ galaxies to provide robust statistical analysis. The redshift distribution has two main characteristics: first, we adopt a bin width of $\Delta{z} = 0.01$ to avoid projection effects that would bias our 2D analysis. The impact of projection effects on the angular homogeneity scale was analyzed in~\cite{Alonso:2013boa}, where it was shown that thin redshift bin width (e.g., $\Delta z=0.01$)  presents minimal bias effects on the homogeneity scale measurements. Second, the redshift spans over an interval of $0.6 < z < 1.0$, so we are able to cover a different period of the Universe compared to previous analyses~\cite{Alonso:2014xca, Goncalves:2017dzs, Avila:2019gdb}. Also, we choose three redshift bins in our work, namely $0.67<z<0.68$, $0.70<z<0.71$, and $0.73<z<0.74$, as they contain the maximum number of LRGs in the DR16 sample (see Fig.\ref{fig:fig2}). Furthermore, as shown in Fig. 2, the number of sources drastically decreases toward higher redshifts, so we restrict our analysis to the intervals mentioned. Otherwise, it would be limited by shot noise.

\section{Method}
\label{sec:Method}

\indent

Our goal is to measure the cosmic homogeneity scale from the angular distribution of galaxies. To do this, we use fractal dimension as our main descriptor of the clustering of point distribution. As pointed out in~\cite{Scrimgeour:2012wt}, we only need the correlation dimension $D_2$ moment to identify the scale of homogeneity, capturing the scaling behavior of the two-point correlation function. Additionally, the correlation dimension $D_2$ is widely accepted as the most reliable measurement of homogeneity when comparing it to its counterparts: the two-point correlation function ($\omega$) and the scaled counts-in-spheres ($\mathcal{N}$) estimators~\cite{Scrimgeour:2012wt, Laurent:2016eqo, Ntelis:2017nrj, Goncalves:2020erb}. Therefore, we only consider the correlation dimension $D_2$ while addressing the cosmic homogeneity scale throughout this work. In what follows, we highlight the main steps toward measuring the cosmic homogeneity scale. 

Firstly, a non-volume-limited sample causes deviation from homogeneity due to survey geometry and completeness. We account for these effects in the same fashion as in~\cite{Scrimgeour:2012wt}, by taking the ratio between our observable by a random catalogue which has the same geometry and completeness as the real data. Moreover, the random catalogue is 50 times denser than the observational data, ensuring statistical fluctuations due to the random points are negligible - see~\cite{eboss20} for details about catalogue creation. 

Second, the conditional probability of finding a galaxy in the element of solid angle $\delta \Omega $ at distance $\theta$ from a randomly chosen galaxy in the ensemble is~\cite{peebles80}
\begin{equation}\label{eq:deltaP}
      \delta P = \bar{\rho}\  \delta \Omega \left[ 1 + \omega(\theta)  \right],
\end{equation}
where $\bar{\rho}$ and $\omega(\theta)$ are the mean density of galaxies and the two-point angular correlation function, respectively. Then, the expected number of neighbors within distance $\theta$ of a galaxy is 
\begin{equation}\label{eq:P}
      P(<\theta) = \bar{\rho} \int d\Omega \left[1 + \omega(\theta)  \right].
\end{equation}
This is precisely the same definition of counts-in-spheres estimators ($N$): the average number of neighboring galaxies in a 2D-sphere of radius $\theta$ around a given galaxy. Therefore, for a random homogeneous distribution (i.e., no clustering, $\omega(\theta)=0$) the counts-in-spheres reads
\begin{equation}\label{eq:NR}
    N_R(<\theta)=2\pi\ \bar{\rho}\ (1-\mathrm{cos}\ \theta)  \, .
\end{equation}
From that, we can define the scaled counts-in-spheres as
\begin{equation}\label{eq:Ncal}
    {\mathcal{N}}(<\theta) \equiv \frac{N(<\theta)}{N_R(<\theta)} =  1 + \frac{1}{1 - \cos\theta} \, \int_{0}^{\theta} \, {\omega}(\theta') \, \sin \theta' \, d\theta'  \, .
\end{equation}
Therefore, we can estimate the above quantity direct from data and random catalogues, using ${\omega}(\theta)$ as the Landy-Szalay two-point angular correlation function estimator\footnote{ In a previous analysis~\cite{Goncalves:2017dzs} we derived and compared measurements of the homogeneity scale from different estimators, such as  the Landy-Szalay and Peebles-Hauser, and found very similar results. In the present analysis, we adopt the Landy-Szalay angular correlation function estimator.}, $\widehat{\omega}_{ls}(\theta)$ ~\cite{Landy:1993yu},

\begin{equation}\label{eq:Ncal}
    \widehat{\mathcal{N}}(<\theta)  =  1 + \frac{1}{1 - \cos\theta} \, \int_{0}^{\theta} \, \widehat{\omega}_{ls}(\theta') \, \sin \theta' \, d\theta'  \, ,
\end{equation}
where 
\begin{equation}\label{eq:w_ls}
\widehat{\omega}_{ls}(\theta) = \frac{DD(\theta) -2DR(\theta) + RR(\theta)}{RR(\theta)},
\end{equation}
and $DD(\theta)$, $DR(\theta)$, and $RR(\theta)$ are numbers of pairs as a function of separation $\theta$, normalized by the total number of pairs, in data-data, data-random and random-random catalogues, respectively. The pair-counting was done with the {\sc TreeCorr} package~\cite{Jarvis:2003wq}.

Finally, as mentioned previously, the correlation dimension $D_2$ provides a more reliable assessment of the scale of homogeneity compared to the scaled counts-in-spheres $\mathcal{N}$ - see~\cite{Scrimgeour:2012wt} for further details. 
We define this quantity following~\cite{Scrimgeour:2012wt, Goncalves:2020erb}, such as
\begin{equation}\label{D2}
    D_2 (\theta) \equiv \frac{d \ln{\cal{N}}(< \theta)}{d \ln{\theta}} + 2 \,.
\end{equation}
Therefore, for a homogeneous distribution, the fractal dimension  $D_2$ is the ambient dimensions, i.e., $D_2(\theta) = 2$, which is valid for small angular scales, as pointed out in~\cite{Goncalves:2017dzs}. We describe the procedure to access the $D_2$ uncertainties in the following.  

In order to obtain the $D_2$ uncertainties, we use two methods to produce new samples. As mentioned earlier, the internal method only uses the data itself. In contrast, the external one needs extra information besides the data to produce new samples. For example, it requires knowledge of the underlying statistics or the physical process which generated the data~\cite{Norberg:2008tg}. In other words, for external methods, one needs to assume a cosmological model to seed simulations, usually called mocks. As we want to perform model-independent measurements of the angular homogeneity scale, our method of choice is internal. However, we also use mocks as our model-dependent baseline, to which we will compare our main results.

Whenever we use the internal method, i.e., 1000 bootstrap resampling technique, or the external method, i.e., 1000 EZmock catalogues~\cite{Zhao:2020bib}, we proceed as follows: We compute $D_2$ using Eq.~(\ref{D2}), and then we get the scale $\theta$ where $D_2=1.98$ using an interpolation procedure. The angular scale where it occurs is hereafter named $\theta_H$, which denotes our \emph{measurements of the angular scale of homogeneity}. We repeat these steps to all 1000 catalogues,  leaving us with a distribution of $\theta_H$, from which we compute the mean and standard deviation as our measurement of the homogeneity scale and its respective uncertainty. Although this $1\%$ criterion consists of an arbitrary choice for determining the homogeneous scale, it has been widely adopted in the literature, so we opt to assume it for consistency with previous works. Additionally, this definition allows for comparison among different surveys, including observational limitations such as survey geometry, besides being easy to be compared with theory, as discussed in~\cite{Scrimgeour:2012wt}.

\section{Results \textbf{and discussion}}
\label{sec:Results}

\indent

\begin{figure}[t!]
	\centering
	\includegraphics[width=0.6\linewidth]{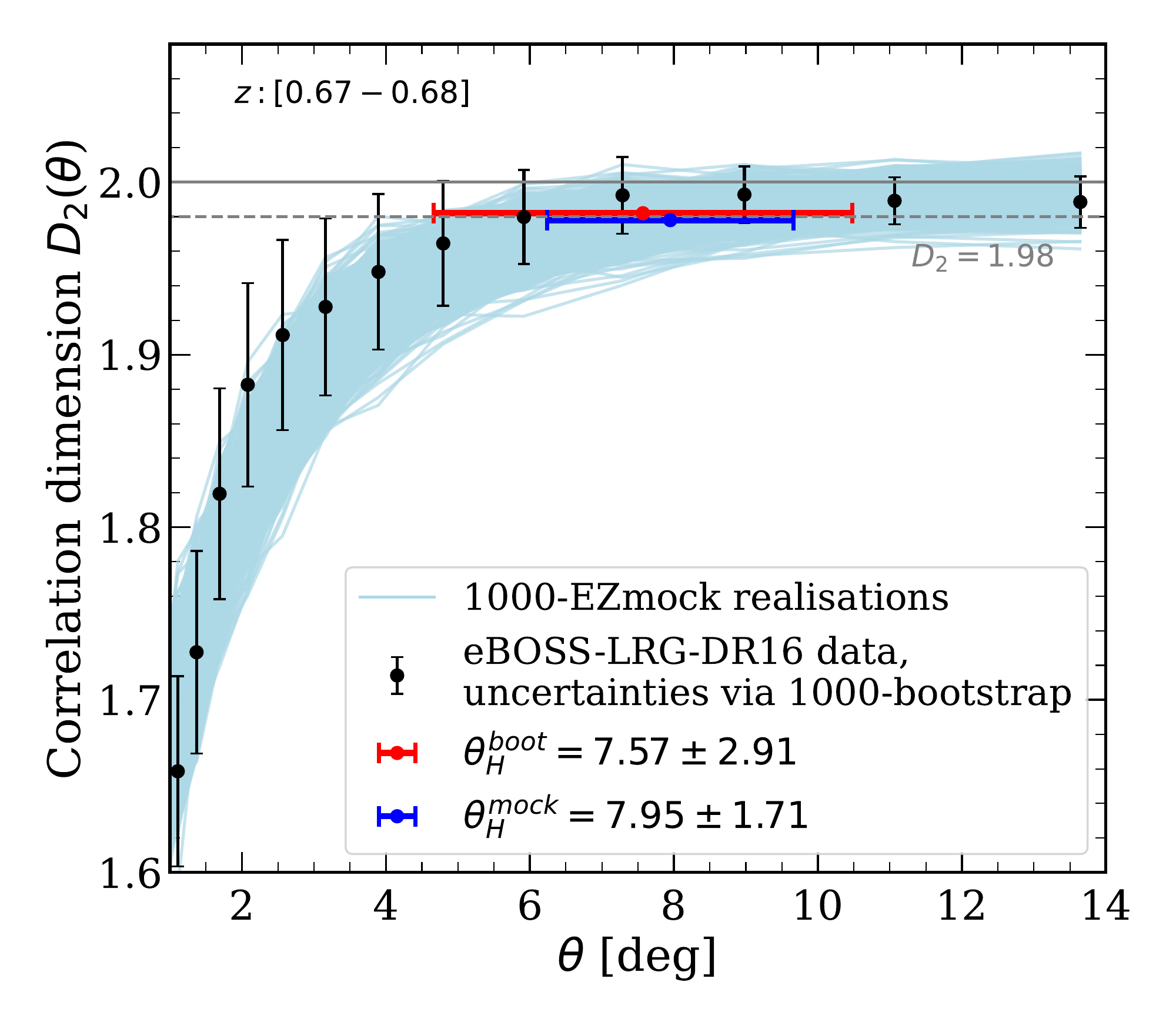}
	\includegraphics[width=0.48\linewidth]{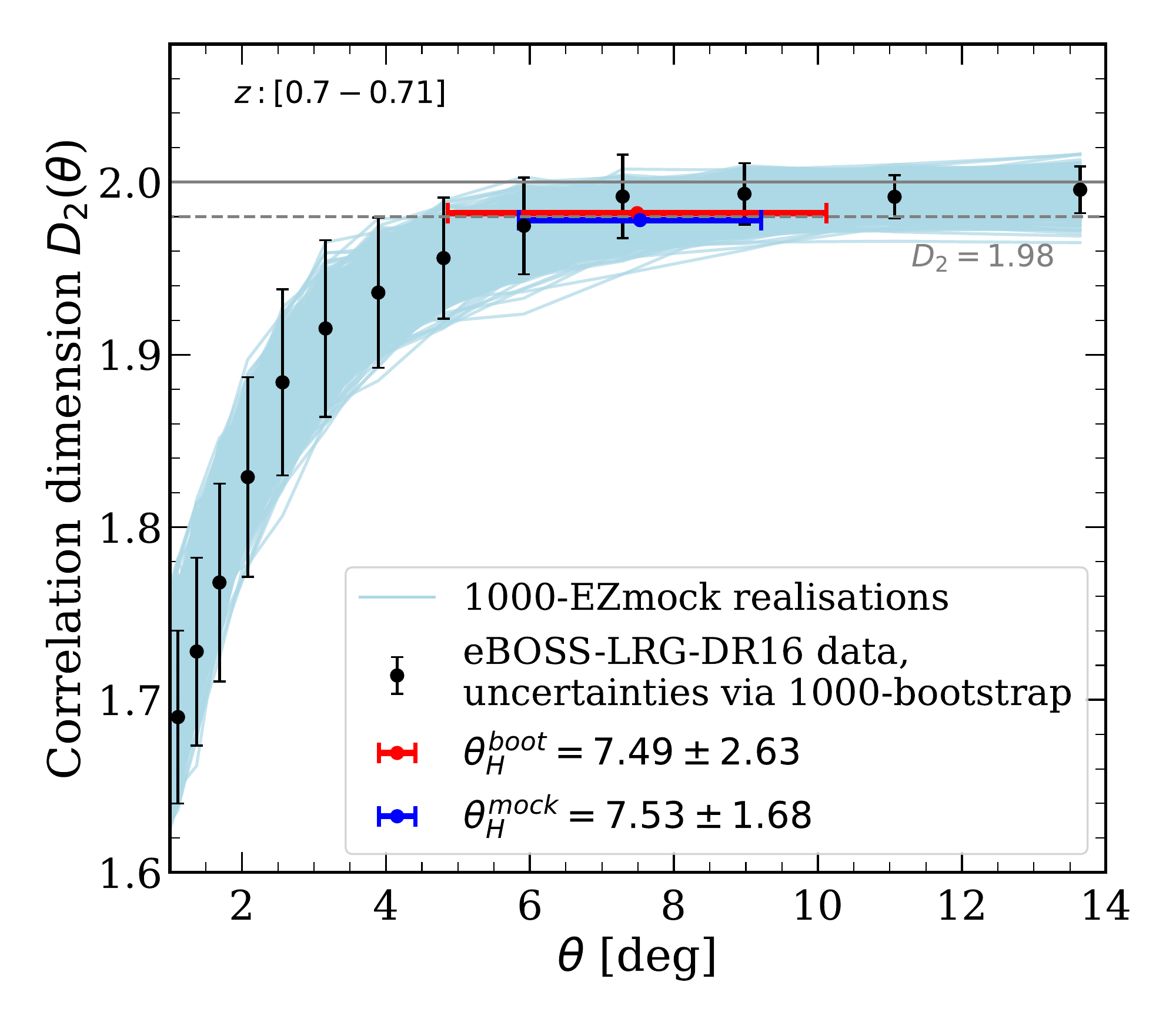}
	\includegraphics[width=0.48\linewidth]{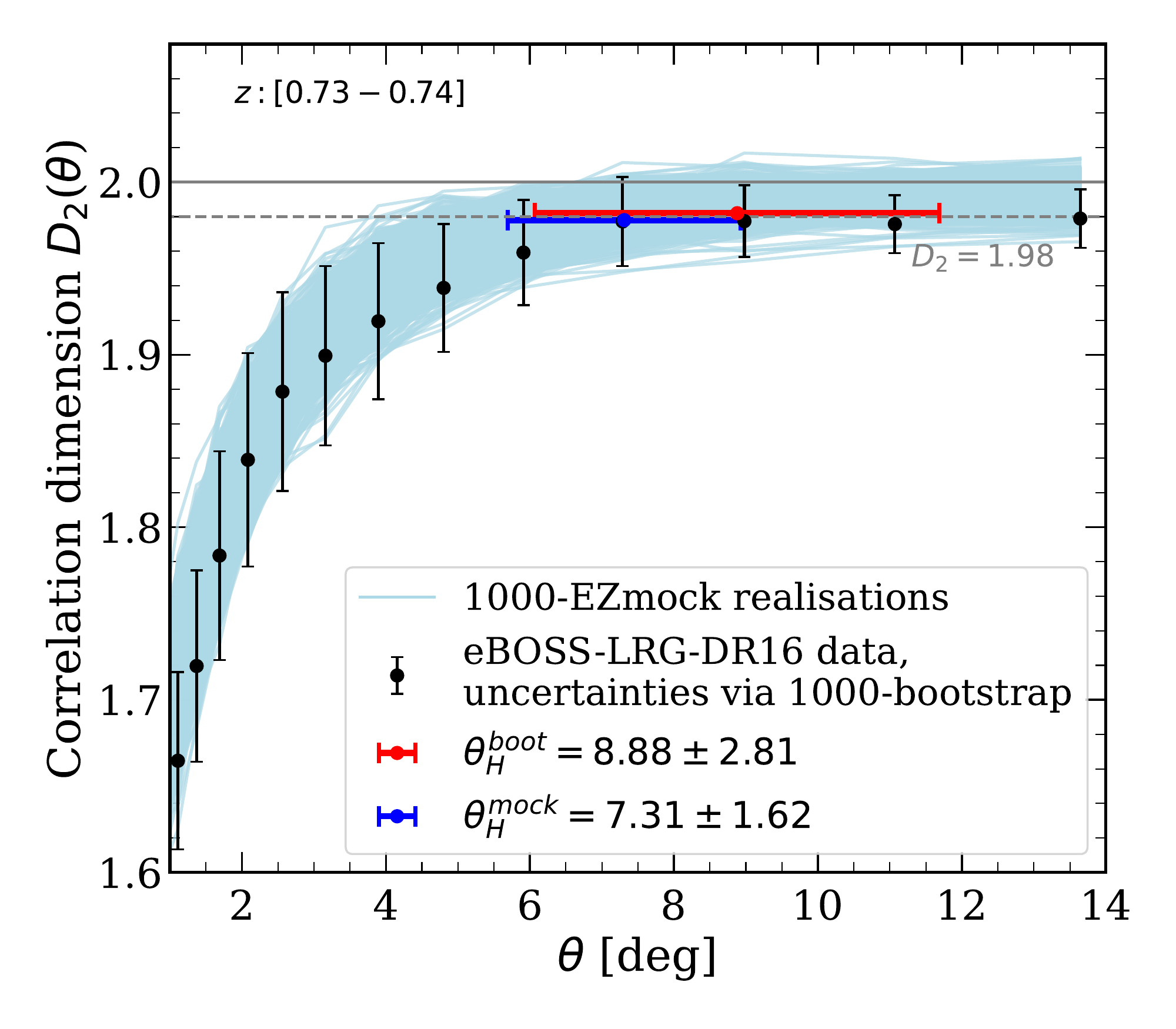}
    \caption{Fractal dimension $D_2(r)$ measurements in each of the three redshift slices. The eBOSS LRG DR16 data $D_2(r)$ measurements are shown as black dots, with their uncertainty given by the internal method, i.e., 1000 bootstrap resamplings of data itself. The lightblue curves are $D_2(r)$ measurements obtained from the external method, namely 1000 realizations of EZmock catalogues. Both methods give consistent angular homogeneity scale measurements, i.e., $\theta_H^{boot}$ agrees with $\theta_H^{mock}$, as depicted in blue and red, respectively.}
    \label{fig:D2_measurements}
\end{figure}

\begin{figure}[th!]
	\centering
	\includegraphics[scale=0.5]{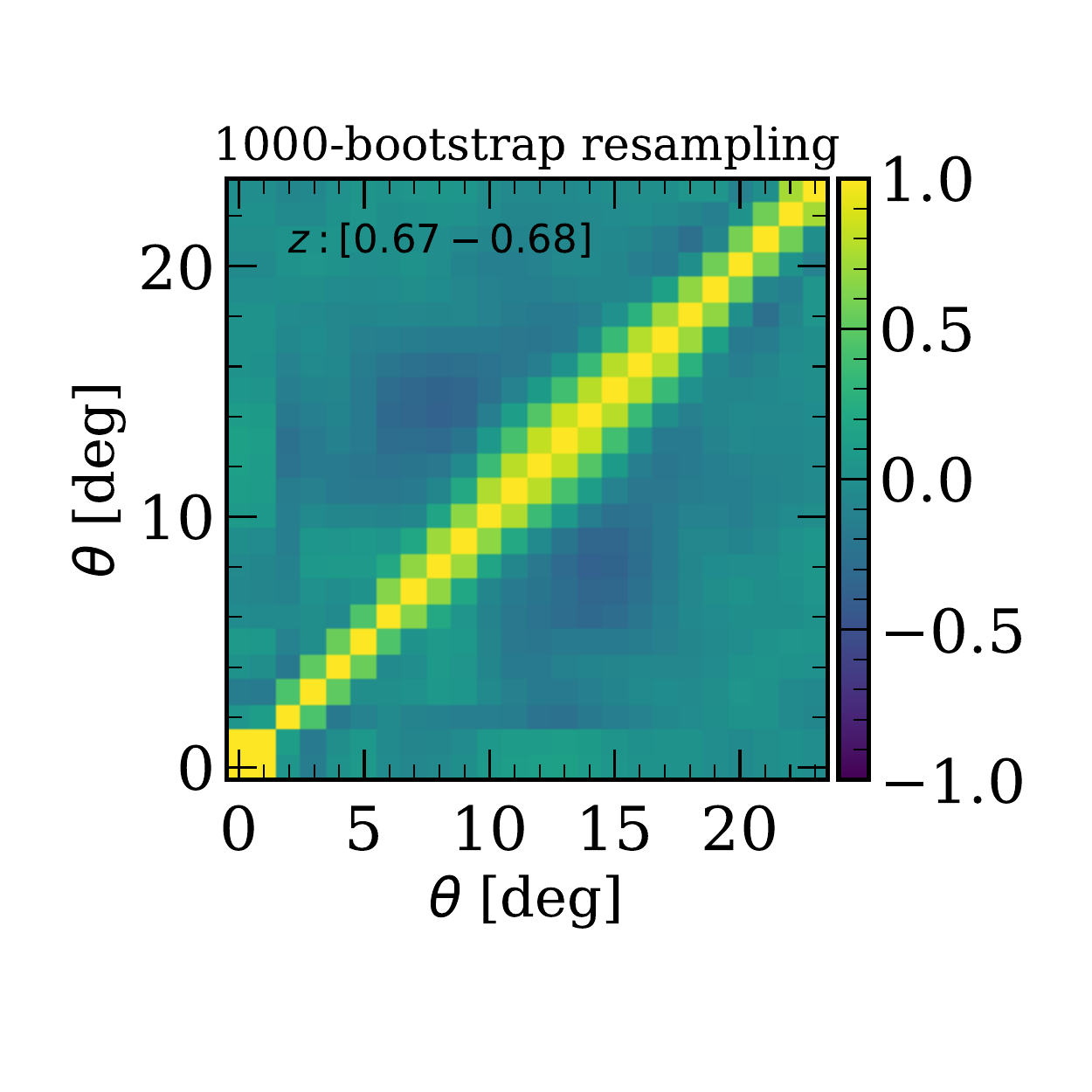}
	\includegraphics[scale=0.5]{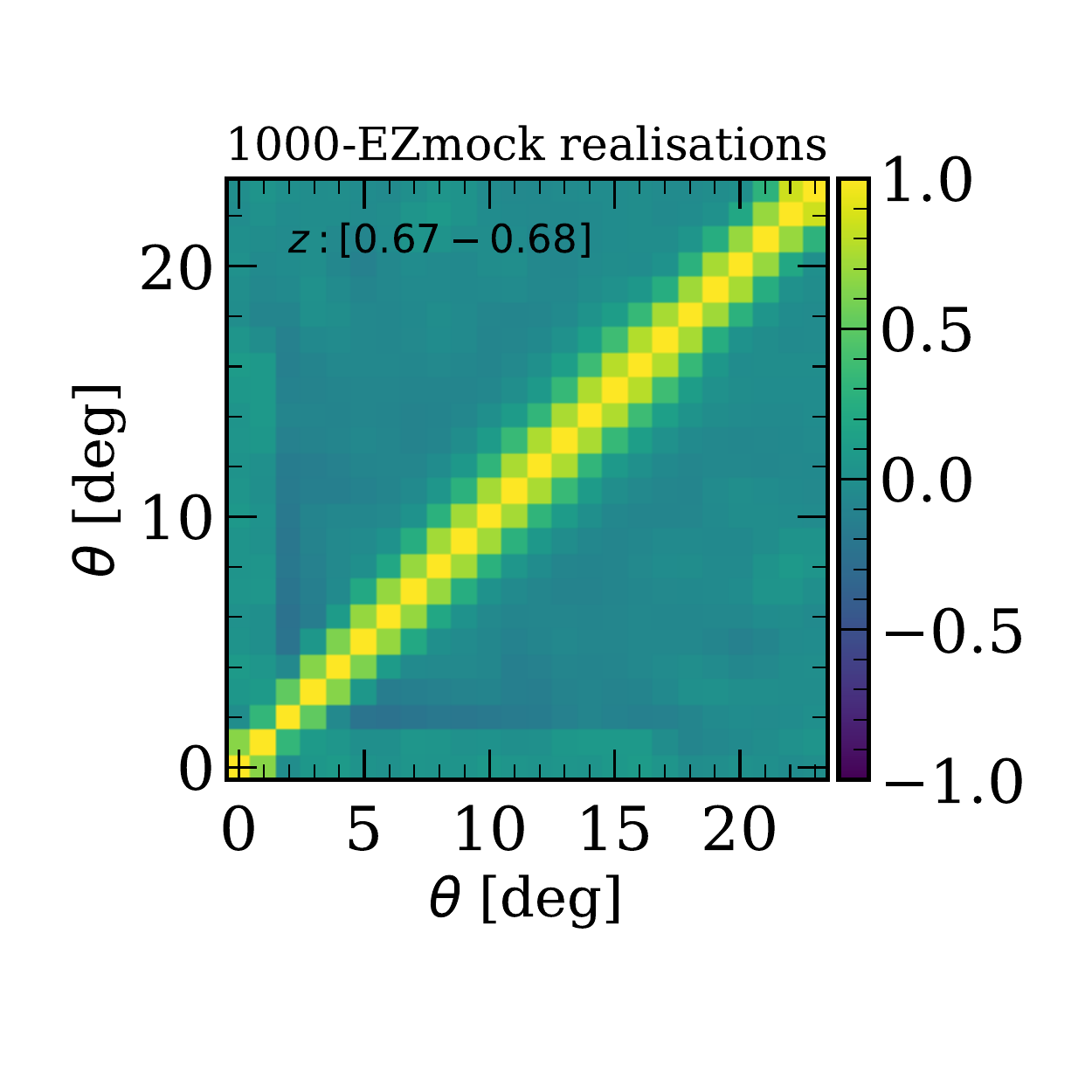}
	\includegraphics[scale=0.5]{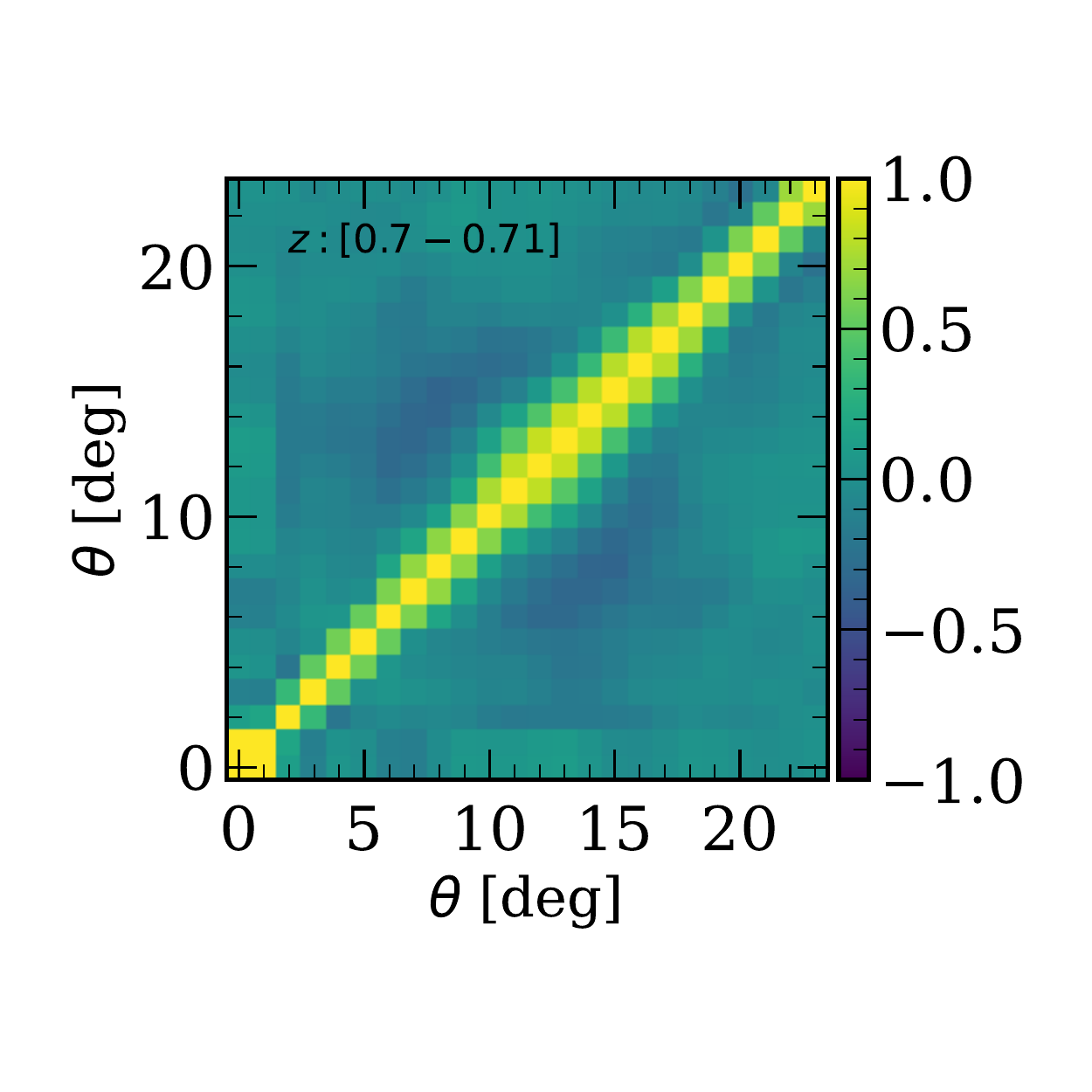}
	\includegraphics[scale=0.5]{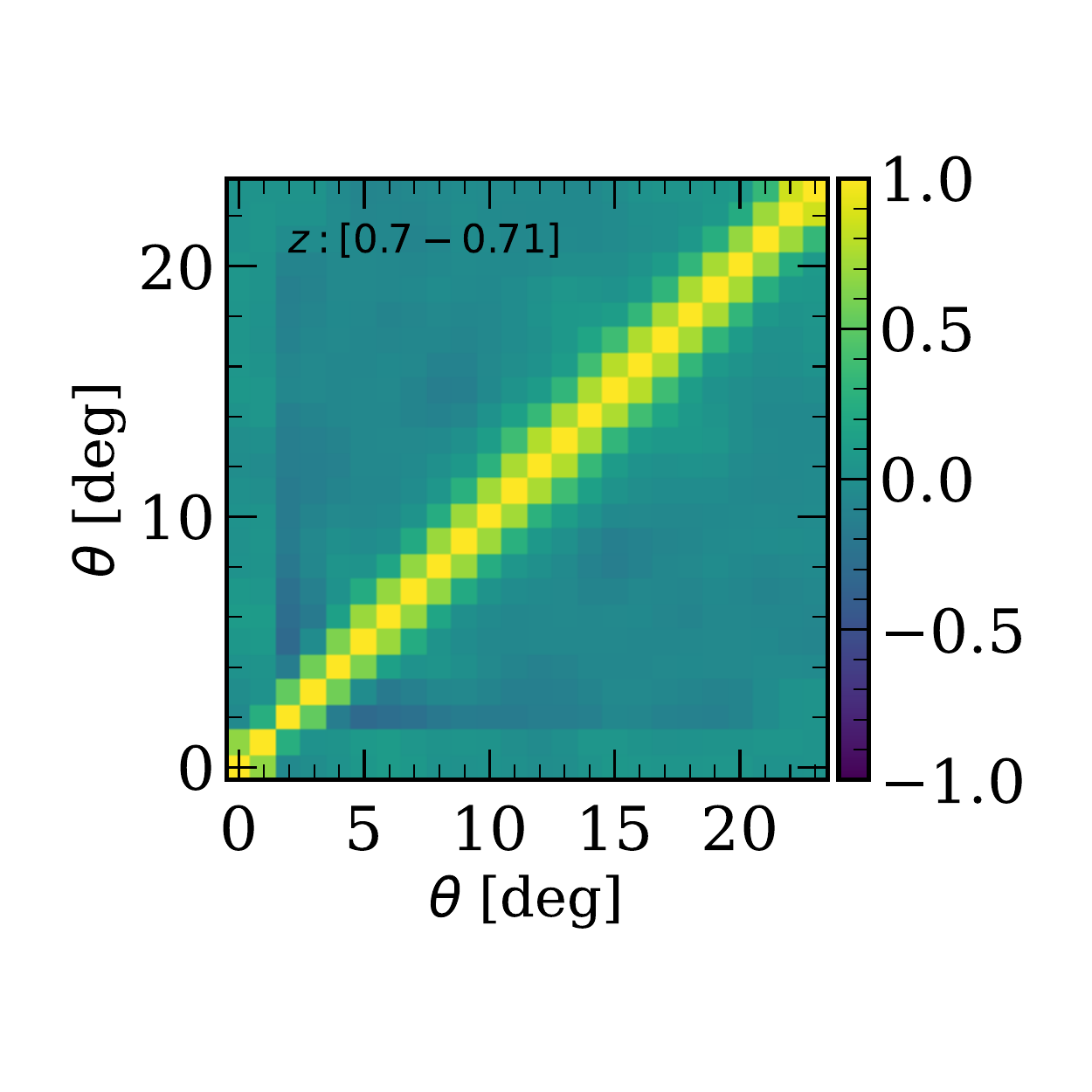}
	\includegraphics[scale=0.5]{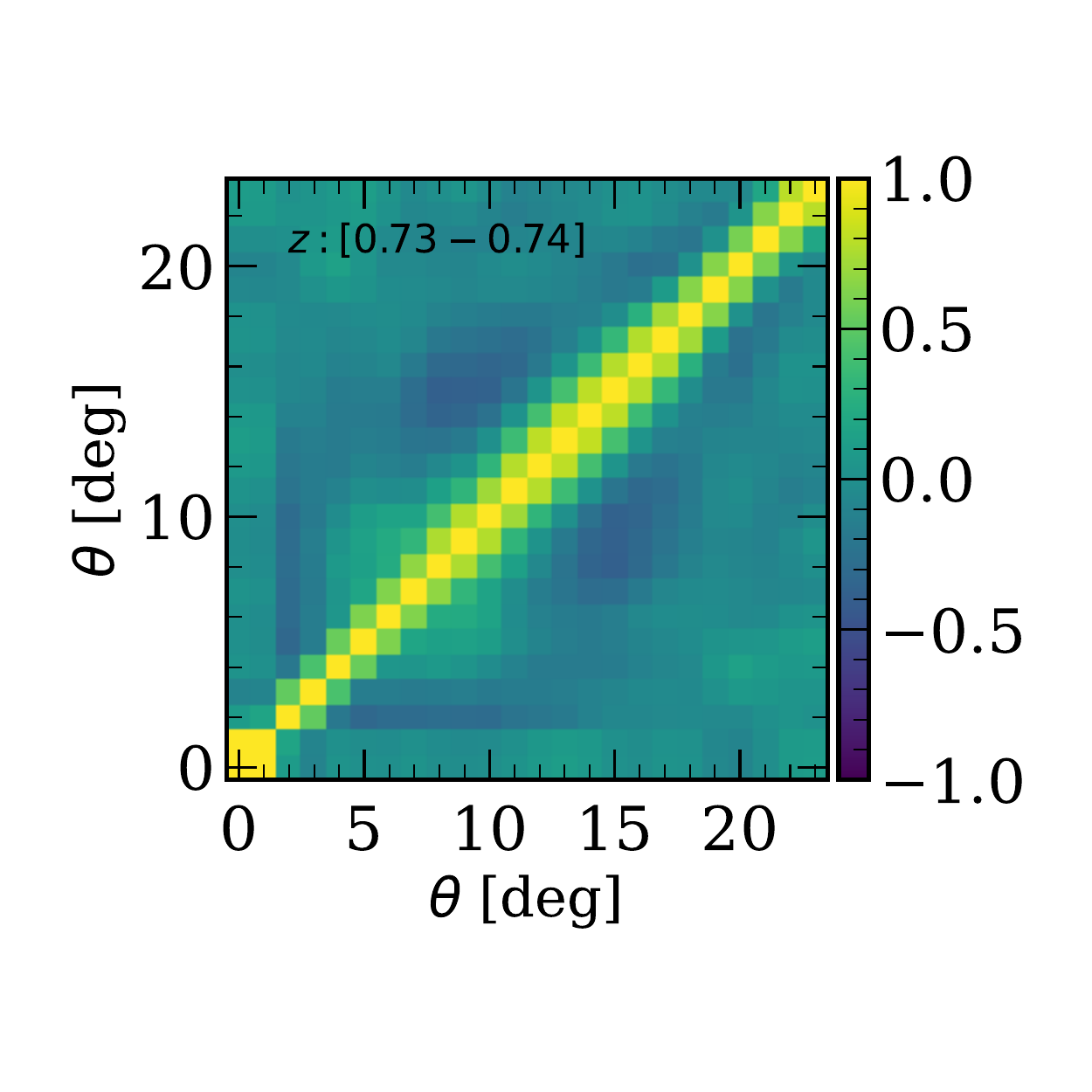}
	\includegraphics[scale=0.5]{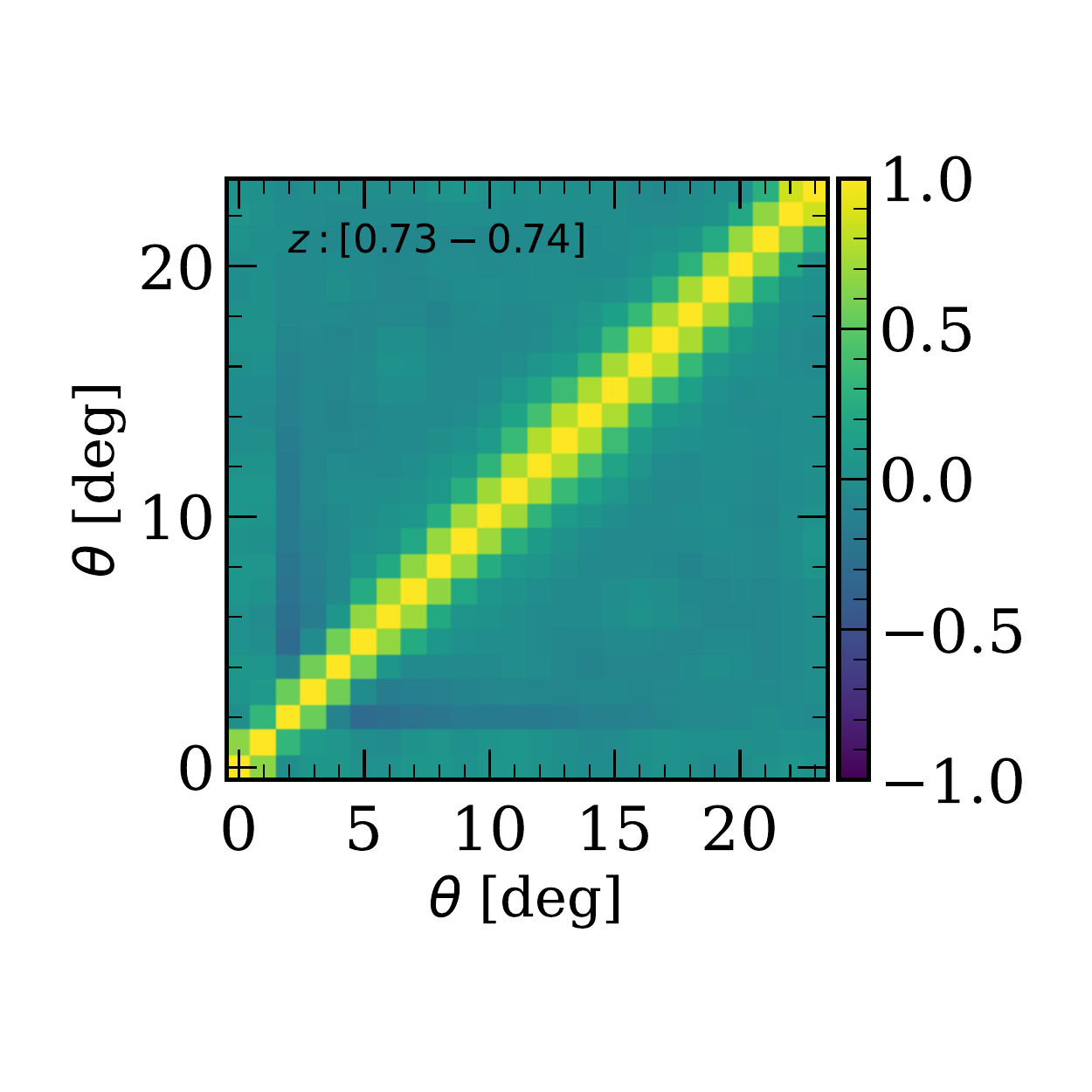}
    \caption{Correlation matrices for $D_2$ in each of the three redshift slices estimated from the 1000 bootstrap resamplings (\textit{left panels}), and the 1000 EZmock realizations (\textit{right panels}).}
    \label{fig:corr_mat}
\end{figure}

The results are shown in Fig~\ref{fig:D2_measurements}. The black points stand for the values of $D_2$ obtained from the observational data at different angular scales whose error bars come from the 1000 bootstrap resampling, as described in the previous section, whereas the lightblue curves denote the $D_2$ values obtained from the 1000 EZmock catalogues.

Since external methods assume that we know the underlying statistical or physical processes that generate the data, we use the internal method as our main measurements to be as model-agnostic as possible. We can see that the $\theta_H^{boot}$ and $\theta_H^{mock}$, respectively shown as red and blue horizontal lines, are in good agreement with each other within $1\sigma$ confidence level, as also shown in Table~\ref{tab:thetah}. This clearly shows consistency between the observed homogeneity scale obtained directly from the data and those obtained considering the underlying $\Lambda$CDM model.

\begin{table}[]
\centering
\begin{tabular}{|c|c|c|}
\hline
 \,\, $\bar{z}$ \,\,& \,\, $\theta_H^{boot}$ (deg) \,\,& \,\, $\theta_H^{mock}$ (deg) \,\, \\
\hline
0.675 & $7.57 \pm 2.91$ & $7.95 \pm 1.71$ \\
0.705 & $7.49 \pm 2.63$ & $7.53 \pm 1.68$ \\
0.735 & $8.88 \pm 2.81$ & $7.31 \pm 1.62$ \\
\hline
\end{tabular}
\caption{Respectively: The redshift interval means, the angular scale of homogeneity using the bootstrap analysis ($\theta_H^{boot}$) and using the 1000 EZmock catalogues ($\theta_H^{mock}$).}
\label{tab:thetah}
\end{table}

Fig.~\ref{fig:corr_mat} shows the correlation matrices for $D_2$ obtained from the internal (left panel) and external (right panel) methods. From this, one can see that the bootstrap analyses capture the same structure as the mock-based method in each of the three redshift bins slices studied. However, bootstrap does present more dispersion in diagonal and off-diagonal elements than mock-based ones. This fact illustrates another benefit of internal methods: any hidden systematic biases that might be missed during mock simulation are by definition taken into account, translating into a wider uncertainty on the scale of homogeneity measurements, as seen in Table~\ref{tab:thetah}. For completeness, in Table~\ref{tab:rh}, we also provide the corresponding 3D physical homogeneity scale ($r_H$) in the canonical flat \lcdm\ model, assuming the Planck18 best-fit~\citep{planck18}. The values are obtained converting the angular scale via $r_H({z}) \equiv d_A({z})\ \theta_H({z})$, where $d_A(z)$ is the angular diameter distance.

\begin{table}[]
\centering
\begin{tabular}{|c|c|c|c|}
\hline
 \,\, $\bar{z}$ \,\,& \,\, $r_H^{boot}$ [Mpc/h] \,\,& \,\, $r_H^{mock}$ [Mpc/h] \,\, \\
\hline
 0.675 & $133.5 \pm 51.3$ & $140.2 \pm 30.2$ \\
 0.705 & $134.4 \pm 47.2$ & $135.1 \pm 30.2 $ \\
 0.735 & $161.9 \pm 51.2$ & $ 133.3 \pm 29.5$ \\
\hline
\end{tabular}
\caption{Respectively: The redshift interval means and the 3D physical homogeneity scale ($r_H$) assuming the Planck18 best-fit~\citep{planck18} in the canonical flat \lcdm\ model for the bootstrap and mock method.}
\label{tab:rh}
\end{table}

\section{Conclusions}
\label{sec:Conclusions}

\indent

A complete reformulation of our description of the large-scale structure of the Universe would have to be pursued if the Cosmological Principle turned out to be invalid. Therefore, it is of fundamental importance to test this underlying assumption of the standard cosmology in light of current observational data (see e.g. \cite{Ellis:2006fy} for a broad discussion). 

In this paper, we probed this assumption by estimating the angular homogeneity scale $\theta_H$ from the recently released LRG sample of the sixteenth data release of the Sloan Digital Sky Survey. The analysis performed is model-independent, as we only used the celestial coordinates of the objects without adopting any fiducial cosmology to convert redshifts into cosmological distances. We divided a sample of 107,500 LRG's -- distributed in the redshift interval of $0.6 < z <1.0$ -- into 3 redshift bins at $\bar z = 0.675, 0.705$ and $0.735$, and suppressed possible correlations between the redshift slices by considering non-contiguous slices. We also adopted a bin width of $\Delta z = 0.01$ to avoid projection effects of objects with large radial separation into the same spherical cap. The uncertainties of the $\theta_H$ estimates were obtained using both mock catalogues and the bootstrap resampling technique, which showed good agreement between them, as displayed in Table II. 

Finally, our results showed a clear scale of cosmic homogeneity in the SDSS LRG DR16 data and {that there is a good agreement between the values of $\theta_H$ extract from the observed data with those obtained from mocks simulations based on the $\Lambda$CDM model}. These results clearly indicate that the standard cosmological hypothesis of large-scale homogeneity is confirmed by the spatial distribution of the LRG's used in our analysis.

\acknowledgments
We acknowledge the use of the Sloan Sky Digital Survey data. It is a pleasure to thank Cheng Zhao for providing early access to the EZmock catalogues used in this work. UA acknowledges financial support from CAPES and Funda\c{c}\~ao de Amparo \`a Pesquisa do Estado do Rio de Janeiro (FAPERJ) Doutorado Nota 10 fellowship at the early stage of this work, and from the Programa de Capacita\c{c}\~ao Institucional PCI/ON. CB acknowledges financial support from PCI/ON at the early stage of this work, and from FAPERJ Postdoc Nota 10 fellowship. JSA is supported by Conselho Nacional de Desenvolvimento Cient\'{\i}fico e Tecnol\'ogico (CNPq 310790/2014-0) and FAPERJ grant 259610 (2021). 

\newpage


\end{document}